\documentclass[aps,prl,reprint,nofootinbib,notitlepage,
amsmath,amssymb,floatfix,longbibliography]{revtex4-2}
\usepackage[T1]{fontenc}
\usepackage{lmodern,graphicx,bm,booktabs,capt-of,mathrsfs,longtable}
\usepackage[normalem]{ulem}
\usepackage[colorlinks=true,linkcolor=blue,citecolor=blue,urlcolor=blue]{hyperref}
\usepackage{bm}
\usepackage{comment}
\newcommand{\dd}{\mathrm d}
\newcommand{\vb}[1]{\bm{#1}}

\newcommand{\etah}{\eta_{c1}}
\def \x {{\bf x}}
\def \y {{\bf y}}
\def \z {{\bf z}}
\def \A {{\bf A}}

\def \D {{\bf D}}
\def \E {{\bf E}}
\def \G {{\bf G}}

\begin{document}
\title{Radiative transitions of hybrid mesons}
\author{Shahriyar Jafarzade}
\email{shah.jzade@pitt.edu}
\affiliation{University of Pittsburgh, Pittsburgh, PA 15260, USA}
\author{Eric S. Swanson}
\email{swansone@pitt.edu}
\affiliation{University of Pittsburgh, Pittsburgh, PA 15260, USA}
\date{\today}

\begin{abstract}
Radiative transitions of hybrid mesons probe nonperturbative gluodynamics, particularly the interplay between quark spin and gluonic excitations, and are of direct relevance to the search for exotic hadrons at Jefferson Lab. We calculate these
transitions across the lowest charmonium-hybrid multiplet
in a constituent gluon model based on Coulomb-gauge QCD in the heavy quark expansion. The results for $Y\to\eta_c\gamma$ and $Y\to \chi_{c0}\gamma$ overlap the width obtained in quenched lattice computations, whereas $\eta_{c1}\to J/\psi\gamma$ is substantially below the
corresponding lattice estimate. 
\end{abstract}
\maketitle
\raggedbottom
\clubpenalty=10000
\widowpenalty=10000

\paragraph*{\textbf{Introduction}.}
The discovery of gluonic excitations is a longstanding goal of hadronic physics. In particular, the properties of 
hybrid mesons provide access to the dynamics of excited gluonic
degrees of freedom in QCD \cite{Meyer:2015eta}. Their identification requires an
understanding of both their spectrum and their decay properties,
motivating theoretical studies alongside experimental spectroscopy
programs such as at GlueX~\cite{GlueX2021}.

Radiative transitions offer an attractive  complementary probe of gluonic structure
because they interrogate how gluonic excitations are absorbed
during photon emission. Understanding radiative transition rates is also central to the experimental program at Hall D at Jefferson Lab, where polarized photons serve as the putative gluonic excitation source.

As an initial study of gluon-photon interactions, we seek to compute radiative transitions of charmonium hybrids to ordinary charmonium states. This is a convenient target because the heavy quarks separate quark and gluonic dynamics to some extent and because quenched lattice calculations exist and can be used to benchmark the model. In particular, the HadSpec collaboration has  obtained widths of
$115(16)\,\mathrm{keV}$ for $\eta_{c1}\to J/\psi\gamma$ and
$42(18)\,\mathrm{keV}$ for a vector hybrid candidate (denoted $Y$) decaying
to $\eta_c\gamma$~\cite{Dudek2009}. More recent work has employed unquenched lattice field theory and extended the method to excited states~\cite{Delaney2024}.

Lattice calculations of excited gluonic potentials and
gluelumps~\cite{Juge1999,FosterMichael1999} provide guidance for
constructing a constituent description of gluonic excitations. Furthermore, calculations of the meson
spectrum identify a lowest hybrid supermultiplet with
$J^{PC}=1^{--},(0,1,2)^{-+}$, consistent with an $S$-wave
quark--antiquark pair coupled to a chromomagnetic
excitation consisting of  an axial
gluonic mode with $(J^{PC})_g=1^{+-}$~\cite{Dudek2011,Liu2012,Cheung2016}. This observation, in turn, meshes well with the Coulomb-gauge Hamiltonian description of QCD, wherein a mean field description of the gluonic vacuum induces a massive axial gluonic quasiparticle \cite{SzczepaniakSwanson2002}. This particle remains transverse, its interactions are completely specified by the Hamiltonian, and because it has an effective mass of approximately 600 MeV, it can be used in a Fock space expansion of hadronic structure. Subsequent applications to the gluelump spectrum \cite{Guo:2007sm}, adiabatic hybrid potentials \cite{Szczepaniak:2005xi,Szczepaniak:2006nx},  hybrid decays \cite{Farina2020,Farina:2023oqk,Farina:2025zhm}, and hybrid mixing \cite{Swanson:2023zlm} are generally quite successful.

Amongst the only previous work on radiative transitions of hybrids we are aware of are a computation using vector meson dominance \cite{Shastry:2023ths} and one that employs the flux tube model of hybrid mesons \cite{Close:2003fz}. In view of the developments just discussed, 
 we  seek to apply the Coulomb gauge formalism to the radiative transitions of hybrids.
An earlier attempt in the same framework 
worked by analogy to the very successful quark model description of mesonic radiative transitions where photons couple to quark and antiquark lines via E1 and M1 operators \cite{Guo2014}. The new element in the hybrid transition then consisted of absorbing the constituent gluon into the instantaneous interaction (a process analogous to the three-gluon interaction of covariant QCD). The resulting integrals were not evaluated, and it was determined that consistent fits to the lattice rates was not 
possible. 

We note however, that power counting in the quark mass indicates that this natural decay mechanism is not accurate. Indeed, the photon coupling is $O(1/m_c)$ while the gluon absorption vertex is $O(1/m_c^2)$\footnote{The power counting emerges because the gluon absorption converts a hybrid to a meson of the same quantum numbers, which necessarily is of opposite total quark spin and  the spin flip operator is suppressed by two powers of mass.}, yielding a highly suppressed amplitude. Alternatively, if the gluon is absorbed in a Z-diagram, while the photon is emitted in a Z-diagram the power counting yields $O(1\cdot 1 \cdot 1/m_c)$, where the final factor comes from the intermediate state. The simplest way to incorporate this mechanism in the computation is to integrate out the Z-diagram, which, of course, amounts to working in the heavy quark effective Hamiltonian approach.

Our approach is therefore to work in the mixed sector of the dual heavy quark effective Hamiltonian of QED and QCD. A leading $O(1/m_c)$ seagull interaction drives the decays, but we also work to $O(1/m_c^2)$ to examine subleading effects. We present results for all possible radiative transitions of the lightest hybrid multiplet  along with several C-violating single-line amplitudes. While the latter are not physically relevant, some have been computed on the lattice, and therefore they supply more points of contact for the current computation.

Our results for $Y \to \eta_c \gamma$, $Y \to \chi_{c0}\gamma$, and the single-line amplitude $\eta_{c1} \to \chi_{c0}\gamma$ agree reasonably well with the lattice field measurements; however, our results for $\eta_{c1} \to J/\psi \gamma$ lie substantially below lattice results. We note that more recent lattice results are in better agreement with our predictions, and regard the situation as unsettled. A lattice computation of any of the other amplitudes we list would help to clarify the situation.

\paragraph*{\textbf{Constituent Hadrons}.}
Meson and hybrid states are constructed by assuming saturation at the lowest available level in the Fock expansion. Thus a rest frame meson is written as

\begin{align}
 |JM[LS]\rangle
 ={}&
 \frac{\delta_{ij}}{\sqrt3}
 \sum_{m,\bar m,M_S,M_L}
 \int\!\frac{\mathrm d^3q}{(2\pi)^3}\,
 \varphi_L(\bm q)\,
 Y_{LM_L}(\hat{\bm q})
 \nonumber\\
 &\times
 \left\langle
 \tfrac12m,\tfrac12\bar m
 \middle|SM_S
 \right\rangle
 \langle SM_S,LM_L|JM\rangle
 \nonumber\\
 &\times
 b^\dagger_{\bm q,i,m}\,
 d^\dagger_{-\bm q,j,\bar m}\,
 |\omega\rangle ,
 \label{eq:meson-state}
\end{align}
where the ket refers to the mean field quark-gluon vacuum and $i$ and $j$ are color indices. 

In a similar way, a rest frame  hybrid  is constructed as \cite{SzczepaniakSwanson2002,Farina2020}

\begin{align}
 \bigl|JM[&LS\ell j_g\xi]\bigr\rangle
 =
 \frac{T^A_{ij}}{2}
 \sum_{\Omega}
 \int\!\frac{d^3q}{(2\pi)^3}
 \frac{d^3k}{(2\pi)^3} \nonumber \\
 &\times
 \Psi_{j_g;\ell }(\bm k,\bm q)
 \sqrt{\frac{2j_g+1}{4\pi}}\,
 D^{j_g*}_{m_g\mu}(\hat{\bm k})\,
 \chi^{(\xi)}_{\mu\lambda} \nonumber \\
 &\times
 \langle\tfrac12m,\tfrac12\bar m|SM_S\rangle \nonumber \\
 &\times
 \langle\ell m_\ell,j_gm_g|LM_L\rangle\,
 \langle SM_S,LM_L|JM\rangle \nonumber \\
 &\times
 b^\dagger_{\bm q-\bm k/2,i,m}\,
 d^\dagger_{-\bm q-\bm k/2,j,\bar m}\,
 a^\dagger_{\bm k,A,\lambda}|\omega\rangle.
 \label{eq:state}
\end{align}
The discrete indices are denoted by 
$\Omega=(m,\bar m,M_S,m_\ell,m_g,M_L,\mu,\lambda)$. This expression couples quark spins to $S$, the quark and gluon angular momenta to $L$, and $S$ and $L$ to $J$. Notice that the gluon polarization couples to the gluon wavefunction via $\chi^{(\xi)}$, which is given by

\begin{align}
\chi^{(+)}_{\mu\lambda} &= \frac{1}{\sqrt{2}}\delta_{\mu,\lambda} ,\nonumber  \\
\chi^{(-)}_{\mu\lambda} &= \lambda\frac{1}{\sqrt{2}}\delta_{\mu,\lambda}.
\end{align}
These expressions correspond to transverse-magnetic (TM) and transverse-electric (TE) hybrids respectively. In this representation the parity and charge conjugation of the hybrid states are given by \cite{Guo2008}

\begin{equation}
P = \xi(-1)^{\ell+j_g+1}, \ \  C = (-1)^{\ell+S+1}.
\end{equation}

Employing TE gluons represents the  model implementation of the desired axial gluonic degrees of freedom. The lowest multiplet of hybrids (denoted $H_1$) is then  obtained by setting  $\ell=0$ and $j_g=L=1$. Setting notation, we call the spin singlet element of this multiplet $Y(1^{--})$, while the
spin-triplet elements are  $H_1(0^{-+})$,
$\eta_{c1}(1^{-+})$, and $H_1(2^{-+})$.

The hybrid wavefunction can be obtained by solving the Tamm-Dancoff equation for the bound state using a novel iterative method that is similar to the Hartree-Fock technique \cite{Swanson:2023zlm}. However, in this exploratory study we simply employ a Gaussian approximation to the wavefunction for all $H_1$ states with scales set by a variational calculation  \cite{Farina2020}. Thus we set 

\begin{equation}
 \Psi_{1; 0}(\bm k,\bm q)
 =
 \mathcal N_H\,k
 \exp\!\left[
 -\frac{q^2}{2\alpha_H^2}
 -\frac{k^2}{2\beta_H^2}
 \right],
 \label{eq:hybrid-radial}
\end{equation}
where the constant $\ell=0$ spherical harmonic is absorbed
into the normalization $\mathcal N_H$.
The widths $\alpha_H$ and $\beta_H$ characterize the
quark-pair and gluon momentum distributions, respectively.
Similarly,  conventional charmonia are represented by
normalized $S$- or $P$-wave Gaussian states, with the
corresponding parameters specified below. These choices permit largely analytic evaluations of the integrals to follow.

\paragraph*{\textbf{The Gluon-Photon Interaction}.}

The nonrelativistic expansion of the Hamiltonian for Coulomb gauge QCD can be obtained by implementing the Foldy-Wouthuysen transformation for field operators    \cite{Szczepaniak:1996tk}. The result closely tracks the expressions from NRQED and NRQCD \cite{Paz:2015uga}:

\begin{align}
H_{QCD} \to H_{FW} &= \int d {\bf x} \left[ m_q \psi^\dagger({\bf x}) \psi({\bf x}) - 
m_{\bar q} \chi^\dagger({\bf x}) \chi({\bf x})\right] \nonumber \\
&+ H_{YM} + H_C + H_1 + H_2 + 
\ldots.
\end{align}
Here $\psi$ annihilates a quark and $\chi$ creates an antiquark. The gluon term is 
\begin{equation}
H_{YM} =  \frac{1}{2}\int d\x \left[ {\cal J}^{-1} \Pi(\x) {\cal J} 
\Pi(\x) + {\bf B}^2(\x) \right],
\end{equation}
where ${\cal J} = \det(\nabla\cdot \mathcal{D})$ is the Faddeev-Popov determinant and 
\begin{equation}
\mathcal{D}^{ab} = \delta^{ab}\nabla + g f^{acb} \G^c,
\end{equation}
is the adjoint covariant derivative. The QCD vector potential is denoted $\G^c$ in the following to distinguish it from the QED vector potential $\A$. The electric field contains transverse and longitudinal components, written as $\vb{\mathcal{E}}^a = - \vb{\Pi}^a + \vb{E}_\parallel^a$ with \cite{Christ:1980ku}

\begin{equation}
\vb{\E}^a_\parallel =  -\nabla G^a_0 - g \nabla \nabla^{-2} f^{abc} \vb{\G}^b\cdot
\nabla G_0^c,
\label{e16}
\end{equation}
and
\begin{equation}
G^a_0(\x) = \frac{1}{g} \int d\y K^{ab}(\x,\y;\G)\rho^b(\y).  
\label{e17}
\end{equation}
The latter introduces a central quantity in the formalism, the instantaneous (non-Abelian) Coulomb kernel
\begin{equation}
K^{ab}(\x,\y;\G) = \langle \x, a| \frac{g}{\nabla\cdot \mathcal{D}} (-\nabla^2) \frac{g}{\nabla\cdot \mathcal{D}}|\y, b\rangle,
\end{equation}
where we employ a braket notation introduced by Lee \cite{Lee1981}. 

The Coulomb kernel builds the Coulomb interaction

\begin{equation}
H_C = \frac{1}{2} \int d\x d\y \, \rho^a(\x) K^{ab}(\x,\y;\G)\rho^b(\y),
\end{equation}
with the color density given by
\begin{equation}
\rho^a(\x) = \psi^\dagger(\x) T^a \psi(\x) + \chi^\dagger(\x)T^a\chi(\x) + g f^{abc} \G^b(\x) \cdot \bm{\Pi}^c(\x).
\end{equation}

The leading correction to the Hamiltonian is given by
\begin{align}
H_1 &= -\frac{1}{2m_c} \int d{\bf x} \psi^\dagger({\bf x}) \left(\D^2 + g \bm{\sigma} \cdot 
{\bf B} \right) \psi({\bf x})  \nonumber \\
&- ( \psi \rightarrow \chi; m_c \rightarrow 
m_{\bar c}).
\end{align}
The subleading correction is 
\begin{align}
H_2 &= -i\frac{g}{8 m_c^2} \int d{\bf x} \psi^\dagger({\bf x}) 
 \bm{\sigma}\cdot ({\bf D} \times {\bf E}- {\bf E}\times \bf{D}) 
\psi({\bf x})  \nonumber \\
&+ ( \psi \rightarrow \chi; m_c \rightarrow m_{\bar c}). 
\end{align}

We require the mixed Hamiltonian that connects photons and gluons. This can be obtained by gauging as 

\begin{equation}
\D  = \nabla -iq \A  -ig \G^a T^a.
\end{equation}
The result is

\begin{align}
 H_{\rm mix}={}&
 \int\!d\x\,
 \psi^\dagger
 \left[
 \frac{qg}{m_c}\A\cdot \G^a
 -\frac{qg}{4m_c^2}\mathcal{S}^a
 \right] T^a\psi  \nonumber\\
 &+
 \int\!d\x\,
 \chi^\dagger
 \left[
 -\frac{qg}{m_c}\A\cdot\G^a
 -\frac{qg}{4m_c^2}\mathcal{S}^a
 \right]T^a\chi,
 \label{eq:mix}\\
 \mathcal S^a={}&
 \vb\sigma\cdot
 \left(
 \vb A\times\vb{\mathcal E}^a
 +\vb G^a\times\vb E
 \right).
 \label{eq:spin}
\end{align}
Notice that the leading piece is a simple seagull interaction that connects a (anti)quark line, a gluon, and a photon at order $1/m_c$. A series of $O(1/m_c^2)$  corrections arising from the spin-orbit interaction are also included in the mixing Hamiltonian\footnote{An effective field theory version of this development contains a Wilson coefficient, $c_S$, which we fix at its tree level value \cite{Manohar1997}.}.

An explicit expression for the Coulomb kernel is required to evaluate the transition amplitudes. Following Ref. \cite{SzczepaniakSwanson2002}, we facilitate this by defining the Faddeev-Popov  amplitude:

\begin{equation}
 F^{ab}(\x,\y;\G)
 =
 \left\langle\x,a\left|
 \frac{-g}{\nabla \cdot \mathcal{D}}
 \right|\y,b\right\rangle .
 \label{eq:F}
\end{equation}
It is useful to extract the vector potential-dependence in this by normal ordering with respect to the constituent vacuum. The result can be written in a generic form as
\begin{align}
 F^{ab}(\x,\y;\G)
 ={}&\delta^{ab}F^{(0)}(\vb x,\vb y)
 \nonumber\\
 &+\int\!d\z\,
 \vb F^{(1)ab,c}(\vb x,\vb y,\vb z)
 \cdot\vb G^c(\vb z)+\ldots .
 \label{eq:Fexp}
\end{align}

We write a similar expansion for the Coulomb kernel as
$K^{ab}=\delta^{ab}K^{(0)}+K^{(1)ab}+\cdots$ and, following Ref. \cite{SzczepaniakSwanson2002}, obtain  an expression for $K^{(1)}$:
\begin{align}
 K^{(1)ab}(\vb x,\vb y)
 \simeq{}&f^{acb}\int\!\dd^3z\,\vb G^c(\vb z)
 \nonumber\\
 &\cdot\Big[
 K^{(0)}(\vb x,\vb z)
 \vb\nabla_z F^{(0)}(\vb z,\vb y)
 \nonumber\\
 &\qquad+
 F^{(0)}(\vb x,\vb z)
 \vb\nabla_z K^{(0)}(\vb z,\vb y)
 \Big].
 \label{eq:K1}
\end{align}
This expression is required in the following analysis because the first term of $\vb{E}_\parallel$ (Eq. \ref{e16}) must be expanded to first order in the field to obtain a gluonic coupling. In contrast, the second term can retain the leading expression for the Coulomb kernel because an explicit gluonic field is present. 

The scalar functions $F^{(0)}$ and $K^{(0)}$ can be obtained by solving coupled Dyson equations using the tree approximation for vertices \cite{SzczepaniakSwanson2002}. The result for $K^{(0)}$ displays a nearly perfect "Coulomb+linear" (Cornell) form that can be compared to lattice results to effect the requisite renormalization. In view of this, one could alternatively use a typical quark model parameterization like 

\begin{align}
 K^{(0)}(r)
 &\approx\frac{\alpha_V}{r}-\frac34br,
 \nonumber\\
 K^{(0)}(p)
 &\approx\frac{4\pi\alpha_V}{p^2}+\frac{6\pi b}{p^4},
 \end{align}
 which we employ in the following. The same formalism yields a  Faddeev-Popov function that can be approximated in simple way as

 \begin{align}
 F^{(0)}(\vb{r}) &\approx \frac{d_0}{\sqrt{2 \pi^3}}\sqrt{\frac{m_g}{r}},
 \nonumber \\
 F^{(0)}(p)
 &\simeq\frac{d_0}{p^2}\sqrt{\frac{m_g}{p}}.
 \end{align}

\begin{figure*}[t]
 \centering
 \includegraphics[width=0.86\textwidth]{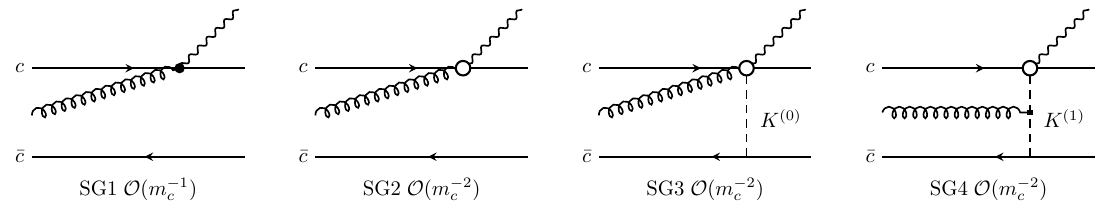}
 \caption{Quark line contributions to $c\bar cg\to c\bar c\gamma$.
 SG1 is the leading seagull term, SG2-SG4 are the $O(1/m_c^2)$ spin-orbit corrections.}
 \label{fig:sgl}
\end{figure*}

The parameters $\alpha_V$ and $b$ are obtained from fits to the charmonium spectrum and are given by 0.546 and 0.1425 GeV$^2$  \cite{Barnes:2005pb}. The parameter $d_0$ emerges from the Dyson equations and is approximately 3.5. The final element of the interaction is $m_g = 600$ MeV which can be interpreted as "gluon mass", and is obtained by fitting a simple form to the  numerical gluon dispersion relation: 

 \begin{equation}
 \omega^2(k)  \approx k^2+m_g^2e^{-k/b_g},
\end{equation}
which was obtained by solving the gluon gap equation for the mean field vacuum. This quantity  feeds into the Dyson equations for the Coulomb and Faddeev-Popov kernels and is important for obtaining the confining solution.
Putting the ingredients together yields the  contributions to the decay amplitude shown in Fig.~\ref{fig:sgl}.

Self energy versions of amplitudes SG3 and SG4 exist. These have two effects: renormalizing the electric and gluonic charges via ultraviolet divergences and adding an infrared-divergent contribution to SG3 and SG4. The latter serve to regulate the (naive) infrared divergences that appear in the momentum space SG3 and SG4 amplitudes due to the strong infrared dependence of $K^{(0)}$ and $F^{(0)}$. Alternatively, working in configuration space obviates infrared divergences due to the meson and hybrid
wavefunctions. We have verified that neglecting the infrared diagrams in this case make no difference (in other words, their infrared behavior solely regulates the momentum space integrals). We have also verified that regulating the naive Fourier transform of $K^{(0)}$ via an exponential or Hadamard regulator yields the same results. Since the ultraviolet behavior is absorbed into parameters, and these are sourced from  other computations, we simply drop the infrared diagrams from further consideration.

Finally, the simple forms of the terms in the mixing Hamiltonian lead to selection rules for the radiative amplitudes:

(i) at leading order $\Delta S=0$;

(ii) at next order $S_H=S_M=0$ amplitudes vanish.

\paragraph*{\textbf{Decay Rates and Amplitudes}.}
We choose the photon momentum along $+\hat{\vb z}$ and
label each amplitude by $h=(m_i,m_f,\nu)$, where the
hadron spin projections refer to this common axis and
$\nu=\pm1$ is the photon helicity.
Angular-momentum conservation requires $m_i=m_f+\nu$.
The four contributions are added before the spin sum:
\begin{align}
 B_h
  =\sum_{r\in\{SG1\ldots SG4\}}B_{r,h},
  \qquad
 \bar B_{\rm tot}^{\,2}
  =\frac{1}{2J_i+1}\sum_h|B_h|^2.
 \label{eq:normalization}
\end{align}
The individual operator norms are defined by
$\bar B_r^{\,2}=(2J_i+1)^{-1}\sum_h|B_{r,h}|^2$ and 
 do not retain the relative phases responsible for
interference.
With dimensionless $B_h$, the amplitude norm $A$ has
units of energy $
 A=2\sqrt{M_iE_f}\,\bar B_{\rm tot}$ where $E_f=\frac{M_i^2+M_f^2}{2M_i}$, and the physical width is
\begin{equation}
 \Gamma=\frac{Q A^2}{8\pi M_i^2},
 \qquad
 Q=\frac{M_i^2-M_f^2}{2M_i}.  
 \label{eq:width}
\end{equation}
Values of $\bar B_{\rm tot}$ inform the amplitude norms which can be compared to lattice results and are reported in Table \ref{tab:physical}.

Physical amplitudes include photon emission from both
the quark and antiquark.
For $C_i=C_f$, these contributions cancel. However, it is useful to examine single-quark-line amplitudes because these have been computed on the lattice.

Parameter choices for the interaction are obtained from Ref. \cite{Barnes:2005pb} and include $\alpha_V = 0.546$, $b = 0.1425$ GeV$^2$, and $m_c = 1.4794$ GeV. Meson widths are obtained by optimizing the overlap with numerical radial wavefunctions and are taken to be 
$\beta_{\eta_c} = 710$ MeV, $\beta_{J/\psi} = 660$ MeV, $\beta_{h_c} = 500$ MeV, and $\beta_{\chi_{cJ}} = 490$ MeV \cite{Farina2020}. The hybrid wavefunction scales are taken from a variational calculation \cite{Close:2005se} and are 
$\alpha_H=415$ MeV and 
$\beta_H=680$ MeV. The parameterization of the constituent gluon dispersion relation uses $b_g = 6.5$ GeV and $m_g = 600$ MeV \cite{SzczepaniakSwanson2002}.
For consistency we choose to set the strong coupling via $g= \sqrt{4\pi \alpha_V}$, which matches the application domain of hadronic interactions at typical binding scales. 

For the hybrid masses, we set 
 $M_Y=4.400$ GeV and $M_{\etah}=4.300$ GeV to match those obtained in the lattice benchmark \cite{Dudek2009}.
 For the remaining partners, we adopt
the lattice-based central values
$M_{H_1(0^{-+})}=4.279$ GeV and
$M_{H_1(2^{-+})}=4.456$ GeV~\cite{Cheung2016}.

\begin{table*}[t]
\centering
\caption{
Radiative transitions of the lowest charmonium hybrids.
The five reduced-amplitude columns display $10^3\bar B$:
individual entries are nonnegative norms, whereas the total
is obtained from the coherent sum of signed amplitudes.
Lattice amplitudes are converted from Ref.~\cite{Dudek2009};
lattice widths are quoted directly with the reported
statistical uncertainties.
}
\label{tab:physical}

\small
\setlength{\tabcolsep}{2.5pt}
\renewcommand{\arraystretch}{1.08}

\begin{tabular*}{\textwidth}
{@{\extracolsep{\fill}}l*{10}{r}@{}}
\toprule
Channel
 & $Q$
 & \multicolumn{5}{c}{$10^3\bar B$}
 & \multicolumn{2}{c}{$A$ [GeV]}
 & \multicolumn{2}{c}{$\Gamma$ [keV]} \\
\cmidrule(lr){3-7}
\cmidrule(lr){8-9}
\cmidrule(lr){10-11}
 & [GeV]
 & SG1 & SG2 & SG3 & SG4 & total
 & Model & LQCD
 & Model & LQCD \\
\midrule

\multicolumn{11}{@{}l}
{\textit{Physical transitions}}\\[2pt]

$Y\to\eta_c\gamma$
 & 1.188
 & 12.612 & 0 & 0 & 0 & 12.612
 & 0.0948 & $0.131(28)$
 & 22 & $42(18)$ \\

$Y\to\chi_{c0}\gamma$
 & 0.875
 & 0 & 8.821 & 0.501 & 1.100 & 9.420
 & 0.0742 & $\lesssim0.105$
 & 10 & $\lesssim20$ \\

$Y\to\chi_{c1}\gamma$
 & 0.799
 & 0 & 9.276 & 0.524 & 1.168 & 9.921
 & 0.0790 & ---
 & 10  & --- \\

$Y\to\chi_{c2}\gamma$
 & 0.763
 & 0 & 11.448 & 0.659 & 1.469 & 12.258
 & 0.0981 & ---
 & 15 & --- \\

\addlinespace[3pt]

$H_1(0^{-+})\to J/\psi\gamma$
 & 1.019
 & 12.737 & 5.207 & 0.307 & 0.653 & 7.184
 & 0.0537 & ---
 & 6  & --- \\

$H_1(0^{-+})\to h_c\gamma$
 & 0.688
 & 0 & 18.412 & 1.097 & 2.447 & 19.762
 & 0.1549 & ---
 & 36 & --- \\

\addlinespace[3pt]

$\eta_{c1}\to J/\psi\gamma$
 & 1.035
 & 12.862 & 3.743 & 0.220 & 0.467 & 10.430
 & 0.0782 & $0.227(16)$
 & 14 & $115(16)$ \\

$\eta_{c1}\to h_c\gamma$
 & 0.705
 & 0 & 8.242 & 0.496 & 1.098 & 8.844
 & 0.0695 & ---
 & 7 & --- \\

\addlinespace[3pt]

$H_1(2^{-+})\to J/\psi\gamma$
 & 1.152
 & 13.689 & 4.943 & 0.279 & 0.592 & 17.349
 & 0.1331 & ---
 & 41 & --- \\

$H_1(2^{-+})\to h_c\gamma$
 & 0.834
 & 0 & 8.577 & 0.487 & 1.078 & 9.168
 & 0.0737 & ---
 & 9 & --- \\

\midrule

\multicolumn{11}{@{}l}
{\textit{Single-quark-line matrix elements}}\\[2pt]

$Y\to J/\psi\gamma^\dagger$
 & 1.110
 & 0 & 2.847 & 0.163 & 0.346 & 3.029
 & 0.0231 & ---
 & --- & --- \\

$Y\to h_c\gamma^\dagger$
 & 0.788
 & 15.702 & 0 & 0 & 0 & 15.702
 & 0.1252 & ---
 & --- & --- \\

\addlinespace[3pt]

$H_1(0^{-+})\to\eta_c\gamma^\dagger$
 & 1.099
 & 0 & 0 & 0 & 0 & 0
 & 0 & ---
 & --- & --- \\

$H_1(0^{-+})\to\chi_{c0}\gamma^\dagger$
 & 0.777
 & 0 & 0 & 0 & 0 & 0
 & 0 & ---
 & --- & --- \\

$H_1(0^{-+})\to\chi_{c1}\gamma^\dagger$
 & 0.699
 & 16.422 & 5.735 & 0.344 & 0.764 & 10.267
 & 0.0804 & ---
 & --- & --- \\

$H_1(0^{-+})\to\chi_{c2}\gamma^\dagger$
 & 0.662
 & 1.998 & 0.662 & 0.034 & 0.081 & 1.288
 & 0.0101 & ---
 & --- & --- \\

\addlinespace[3pt]

$\eta_{c1}\to\eta_c\gamma^\dagger$
 & 1.115
 & 0 & 1.833 & 0.109 & 0.228 & 1.953
 & 0.0145 & $\simeq0.067(9)$
 & --- & --- \\

$\eta_{c1}\to\chi_{c0}\gamma^\dagger$
 & 0.794
 & 8.808 & 3.785 & 0.214 & 0.478 & 12.857
 & 0.0998 & $\simeq0.060(10)$
 & --- & --- \\

$\eta_{c1}\to\chi_{c1}\gamma^\dagger$
 & 0.717
 & 8.187 & 0.555 & 0.027 & 0.066 & 8.557
 & 0.0672 & ---
 & --- & --- \\

$\eta_{c1}\to\chi_{c2}\gamma^\dagger$
 & 0.679
 & 10.902 & 8.078 & 0.482 & 1.076 & 2.659
 & 0.0210 & ---
 & --- & --- \\

\addlinespace[3pt]

$H_1(2^{-+})\to\eta_c\gamma^\dagger$
 & 1.229
 & 0 & 1.569 & 0.090 & 0.188 & 1.667
 & 0.0126 & ---
 & --- & --- \\

$H_1(2^{-+})\to\chi_{c0}\gamma^\dagger$
 & 0.919
 & 1.562 & 0 & 0 & 0 & 1.562
 & 0.0124 & ---
 & --- & --- \\

$H_1(2^{-+})\to\chi_{c1}\gamma^\dagger$
 & 0.845
 & 7.577 & 5.828 & 0.330 & 0.729 & 13.704
 & 0.1099 & ---
 & --- & --- \\

$H_1(2^{-+})\to\chi_{c2}\gamma^\dagger$
 & 0.809
 & 13.151 & 0.846 & 0.040 & 0.096 & 13.078
 & 0.1054 & ---
 & --- & --- \\

\bottomrule
\end{tabular*}
\end{table*}

Table~\ref{tab:physical} reports amplitudes and decay rates for ten physical transitions and 14
single-line amplitudes, along with amplitudes and rates from the lattice field theory computations of Ref. \cite{Dudek2009}. 

We note immediately that the leading seagull average amplitudes, $\bar B$, are nearly the same for all the physical processes. This is because the vertex is very simple, a global spin-averaged spatial wavefunction is used for the hybrids, the photon energies are comparable, and the mesonic wavefunctions are similar. Thus, one obtains a generic prediction

\begin{equation}
\Gamma(H \to M\gamma) \sim (10-20) \textrm{ keV} + O(\Lambda_{QCD}/m_c).
\end{equation}
In comparison, the lattice computation gives $\Gamma(Y \to \eta_c\gamma) = 42(18)$ KeV and $\Gamma(\eta_{c1} \to J/\psi \gamma) = 115(16)$ KeV.
Although the order of magnitude is correct, the factor of three between these rates is difficult to understand with the current formalism. Possible causes are substantial shifts in the hybrid structure, or subleading effects. These are zero for $Y\to \eta_c \gamma$, and it is therefore gratifying that this prediction is reasonably close to the lattice result. Computing the spin-orbit contributions for $\eta_{c1} \to J/\psi\gamma$ reveals amplitudes of reduced, but comparable, size to the leading seagull. It is therefore unfortunate that the phases involved tend to lead to cancellation in this channel and drop the model prediction to about one half of the $Y$ decay rate.

Predictions for ten accessible radiative transitions are presented. Eight of these have not been computed in lattice field theory; the predicted amplitudes are similar; however, one can expect that $\Gamma(H_1(2^{-+})\to J/\psi \gamma) \approx 4 \cdot\Gamma(H_1(2^{-+}) \to h_c\gamma)$ and that the widths for $Y\to \chi_{cJ}\gamma$ should be comparable to each other. 

We note that a more recent unquenched lattice computation of charmonium decay rates has been made, yielding results of 30 and 105 keV, somewhat closer to our predictions \cite{Delaney2023}. Also, our value for $\Gamma(Y \to \chi_{c0}\gamma)$ satisfies the lattice upper limit for this reaction.

Fourteen single line amplitudes are reported in Table \ref{tab:physical}; of these, $\eta_{c1}\to \chi_{c0}\gamma$ agrees reasonably well with the lattice result, while $\eta_{c1} \to \eta_c\gamma$ lies well below it. Interesting patterns are apparent for the $H_1(S=1)$ hybrid single line transitions to the $\chi_{cJ}$ states,  with the table showing unique $J$-dependence among the three states. 
We anticipate that exploring these ratios in lattice field theory would be a useful diagnostic for any putative hybrid models.

\paragraph*{\textbf{Parameter Sensitivity and Lattice Error Budget}.}
We have estimated sensitivity of the model results to parameter variation by directly recomputing. We find that a 10\% variation in the hybrid quark wavefunction scale, $\alpha_H$ leads to approximately 10\% variation in the widths. Similarly, a 10\% variation in the gluonic wavefunction scale, $\beta_H$ gives a 20\% variation in the predicted widths. The latter also applies to  10\% variations in the mesonic wavefunction scales. 

\begin{figure}[t]
\centering
\includegraphics[width=0.9 \columnwidth]{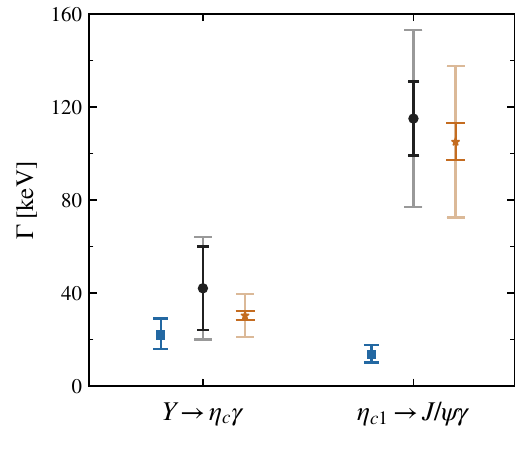}
\caption{Radiative widths: model predictions (blue squares),
quenched LQCD (black circles)~\cite{Dudek2009}, and
$N_f=2+1$ results (orange stars)~\cite{Delaney2023}.
Blue bars span individual $\pm10\%$ wavefunction-width variations.
The orange $Y\to\eta_c\gamma$ point is our reconstruction,
$30.3\pm1.8$~keV, with approximate propagated uncertainty
including fit-form spread.
Other inner bars show reported statistical errors; lighter
outer bars add an estimated systematic  uncertainty.}
\label{fig:width-sensitivity}
\end{figure}

The errors reported for the lattice computations in Table \ref{tab:physical} are statistical; the authors cite a number of sources of systematic errors that "can be addressed in future calculations". Since these errors can be substantial we have chosen to estimate them by comparing their predictions for known meson radiative transitions to experiment \cite{ParticleDataGroup:2026mpi}. The results indicate an approximate additional uncertainty of 30\%. This estimate is included in Fig. \ref{fig:width-sensitivity}, which also shows the newer lattice results of Delaney \cite{Delaney2023}.

\paragraph*{\textbf{Hybrid to Hybrid Radiative Transitions}.}

Hybrid to hybrid radiative transitions may be of interest to some future experiment or as an additional test for lattice field theory. These are easily computed in the formalism and map directly to traditional quark model computations for mesonic transitions since the constituent gluon acts as a spectator. 

Table~\ref{tab:hybrid-hybrid} lists the spectator estimates
and photon momenta for the kinematically allowed channels amongst the low lying $H_1$ multiplet. Note that in the long-wavelength limit, these M1 widths scale as
$Q^3$, making the predictions particularly sensitive
to the poorly known mass splittings within the multiplet.


\begin{table}[t]
\centering
\caption{Radiative transitions within the lowest
charmonium-hybrid multiplet in the spectator approximation.}
\label{tab:hybrid-hybrid}
\begin{tabular}{lrr}
\toprule
Transition & $Q$ [MeV] & $\Gamma$ [keV] \\
\midrule
$Y\to H_1(0^{-+})\gamma$       & 119.3 & 0.463 \\
$Y\to\eta_{c1}\gamma$  &  98.9 & 0.605 \\
$H_1(2^{-+})\to Y\gamma$       &  55.6 & 0.228 \\
\bottomrule
\end{tabular}
\end{table}

\paragraph*{\textbf{Discussion}.}
We have calculated radiative transitions of the lowest
charmonium-hybrid multiplet in the Coulomb-gauge
constituent-gluon model, including the leading 
seagull contact and three spin-dependent contributions.
The seagull contributes to spin-preserving transitions,
while spin-changing channels directly probe the 
subleading interactions. 

Our results for $Y\to \eta_c\gamma$ agree with lattice computations within rather substantial errors. In contrast the rate for $\eta_{c1}\to J/\psi \gamma$ lies well below the lattice result.

We offer predictions for ten physical and an additional 14 single line amplitudes for comparison to possible future lattice computations. Of note are the distinct decay patterns of the $0^{-+}$ and $2^{-+}$ hybrid partner states into  $J/\psi\gamma$ and $h_c\gamma$. In a similar way, the single line amplitudes for $0^{-+}$, $1^{-+}$, and $2^{-+}$ to $\chi_{cJ}$ offer contrasting and distinctive decay patterns that would be useful for testing and diagnosis. Certainly, 
more data would serve as a stringent test of the utility of the constituent gluon picture of low energy gluodynamics and would assist in building robust methodology for hybrid discovery. 

It will be of interest to extend the methods here to light hybrids. One approach would be to adopt the traditional quark model approach of replicating heavy quark dynamics with light quarks while allowing parameter variation to absorb model deficiencies. Success would permit computing radiative transitions that are relevant to the GlueX experiment at Jefferson Lab.

\paragraph*{\textbf{AI Usage}.} The authors used GPT-6 Astra (OpenAI) to assist in checking the numerical evaluation of the SG3 and SG4 integrals. Claude Opus 5.5 was used for grammar and spell checking. The authors take full responsibility for the results and conclusions presented in this work.

\begin{acknowledgments}
The authors acknowledge support from the U.S.\ Department of Energy under
Grant No.\ DE-SC0019232. This work contributes to the goals of the US DOE
ExoHad Topical Collaboration, Contract No.\ DE-SC0023598.
\end{acknowledgments}

\clearpage
\onecolumngrid
 
\providecommand{\vb}[1]{\boldsymbol{#1}}
\providecommand{\dd}{\mathop{}\!\mathrm{d}}
\setcounter{section}{0}
\setcounter{equation}{0}
\renewcommand{\thesection}{S\arabic{section}}
\renewcommand{\theequation}{S\arabic{equation}}
\ifdefined\theHsection
  \renewcommand{\theHsection}{supp.\arabic{section}}
\fi
\ifdefined\theHequation
  \renewcommand{\theHequation}{supp.\arabic{equation}}
\fi

\section*{End Matter}

Explicit expression for the transition matrix elements are as follows.
We denote $\vb\epsilon_g=\vb\epsilon(\vb k,\lambda)$, 
$\vb e_\nu^*=\vb\epsilon^*(\vb Q,\nu)$, and $\Delta_\pm=\frac{\pm\vb k-\vb Q}{2}\mp\vb P$.
Single-quark-line matrix elements are
\begin{align}
 \mathscr M_{{SG1},h}^{(q)}
 ={}&\frac{2q_cg}{\sqrt3\,m_c}\,
 \mathcal P_h\int_{\vb k,\vb q,\vb P}
 \psi_H(\vb k,\vb q)\varphi_f^*(\vb q+\boldsymbol\Delta_+)
\frac{\vb\epsilon_g\cdot\vb e_\nu^*}
 {\sqrt{2\omega(k)}\sqrt{2Q}}\,\delta(\vb P),
 \label{eq:supp-full-SG}\\[3pt]
 \mathscr M_{SG2,h}^{(q)}
 ={}&-\frac{2i q_cg}{4\sqrt3\,m_c^2}\,
 \mathcal P_h\int_{\vb k,\vb q,\vb P}
 \psi_H(\vb k,\vb q)\varphi_f^*(\vb q+\boldsymbol\Delta_+)
 \frac{\boldsymbol\sigma_c\cdot
 (\vb e_\nu^*\times\vb\epsilon_g)}
 {\sqrt{2\omega(k)}\sqrt{2Q}}\,[\omega(k)+Q]\,\delta(\vb P),
 \label{eq:supp-full-SGL2}\\[3pt]
 \mathscr M_{SG3,h}^{(q)}
 ={}&-\frac{i\sqrt3 q_cg}{4m_c^2}\,
 \mathcal P_h\int_{\vb k,\vb q,\vb P}
 \psi_H(\vb k,\vb q)\varphi_f^*(\vb q + \Delta_-)
  \frac{
 [\boldsymbol\sigma_c\cdot(\vb e_\nu^*\times\vb P)]
 (\vb\epsilon_g\cdot\vb P)}
 {\sqrt{2\omega(k)}\sqrt{2Q}}\, K(|\vb P - \vb k|)\frac{1}{P^2},
 \label{eq:supp-full-SGL3}\\[3pt]
 \mathscr M_{SG4,h}^{(q)}
 ={}&+\frac{i\sqrt3 q_c}{4m_c^2}\,
 \mathcal P_h\int_{\vb k,\vb q,\vb P}
 \psi_H(\vb k,\vb q)\varphi_f^*(\vb q + \Delta_-)
 \frac{
[\boldsymbol\sigma_c\cdot(\vb e_\nu^*\times\vb P)]
 (\vb\epsilon_g\cdot\vb P)}
 {\sqrt{2\omega(k)}\sqrt{2Q}}
 \big[F(P)K(|\vb P-\vb k|)
              +K(P)F(|\vb P-\vb k|)\big].
 \label{eq:supp-full-SGL4}
\end{align}

Lattice quantities are converted to physical units as follows. 
For a
vector-to-pseudoscalar form factor $\widehat V(0)$,
\begin{equation}
 A_{VP}=\mathcal L q_c\sqrt{\frac23}
 \frac{2M_iQ}{M_i+M_f}|\widehat V(0)|.
 \label{eq:latVP}
\end{equation}
Here $\mathcal L$ accounts for the current attachments:
$\mathcal L=2$ for physical transitions with $C_i=-C_f$,
where the quark and antiquark contributions add coherently,
and $\mathcal L=1$ for the single-quark-line matrix elements. The latter are quoted for comparison
only; their full electromagnetic amplitudes vanish because
the two attachments cancel when $C_i=C_f$.
For the dimensionful transverse multipoles in the convention
of Ref.~\cite{Dudek2009}, the spin-averaged amplitude norm is
\begin{equation}
 A=\mathcal L q_c
 \left[
 \frac{2}{2J_i+1}
 \sum_{k\,\mathrm{transverse}}
 |\widehat F_k(0)|^2
 \right]^{1/2}.
 \label{eq:latmultipole}
\end{equation}
The sum includes the electric and magnetic multipoles
contributing to real-photon emission.
Values quoted in temporal lattice units,
$a_t\widehat F_k$, are converted to GeV using
$a_t^{-1}=6.05$~GeV. 

\twocolumngrid

\bibliography{main}

\end{document}